\documentclass[smallextended]{svjour3}       
\usepackage{pdflscape}
\usepackage{afterpage}
\usepackage{capt-of}\usepackage{rotating}
\usepackage{graphicx}
\graphicspath {{./figures/}}
\usepackage{enumitem}
\usepackage{array}
\usepackage{url}
\usepackage{booktabs}
\usepackage{cite}
\usepackage{subcaption}
\usepackage{multicol}
\usepackage{multibib}
\usepackage{color}
\usepackage{ifthen}
\usepackage{mdframed}
\usepackage{hyperref}
\mdfdefinestyle{boxstyle}{rightline=true,innerleftmargin=4,innerrightmargin=5,linewidth=1pt}
\mdfsetup{skipabove=5pt,skipbelow=5pt}

\newcites{icse}
{References to papers within the study sample}    

\newcommand{\replicationUrl}{\href{http://doi.org/10.5281/zenodo.3813878}{DOI: 10.5281/zenodo.3813878}} 
\newcolumntype{P}[1]{>{\centering\arraybackslash}p{#1}}

\journalname{Empirical Software Engineering}
\begin{document}

\title{The Who, What, How of Software Engineering Research: A Socio-Technical Framework}
\titlerunning{The Who, What, How of Software Engineering Research}

\author{Margaret-Anne Storey \and 
 		Neil A. Ernst \and
        Courtney Williams \and
        Eirini Kalliamvakou
}

\authorrunning{Storey et al.} 
\institute{Margaret-Anne Storey \and Neil A. Ernst \and Courtney Williams \and Eirini Kalliamvakou \at
            University of Victoria, Canada\\
            \email{mstorey@uvic.ca, nernst@uvic.ca, courtneyelwilliams@gmail.com, ikaliam@uvic.ca}
}

\date{Received: date / Accepted: date}

\maketitle

\begin{abstract}

Software engineering is a socio-technical endeavor, and while many of our contributions focus on technical aspects, human stakeholders such as software developers are directly affected by and can benefit from our research and tool innovations. 
In this paper, we question how much of our research addresses human and social issues, and explore how much we study human and social aspects in our research designs. 
To answer these questions, we developed a socio-technical research framework to capture the main beneficiary of a research study (the \textit{who}), the main type of research contribution produced (the \textit{what}), and the research strategies used in the study (\textit{how} we methodologically approach delivering relevant results given the \textit{who} and \textit{what} of our studies). 
We used this Who-What-How framework to analyze 151 papers from two well-cited publishing venues---the main technical track at the International Conference on Software Engineering, and the Empirical Software Engineering Journal by Springer---to assess how much this published research explicitly considers human aspects.  
We find that although a majority of these papers claim the contained research should benefit human stakeholders, most focus on technical contributions without engaging humans in their studies.

Although our analysis is scoped to two venues, our results suggest a need for more diversification and triangulation of research strategies. In particular, there is a need for strategies that aim at a deeper understanding of human and social aspects of software development practice to balance the design and evaluation of technical innovations. 
We recommend that the framework should be used in the design of future studies in order to nudge software engineering research towards explicitly including human and social concerns in their designs, and to improve the relevance of our research for human stakeholders.

\end{abstract}

\section{Introduction}
\label{cha:introduction}

Nowadays we recognize software engineering as a socio-technical endeavor~\cite{Whitworth2011}, and we increasingly see social aspects as a critical part of the software engineering practice and research landscape~\cite{FTAS08}. What is more, while we may expect that many of our contributions are purely technical, somewhere, at some time, a software developer may be affected by our work. It is crucial to account for the social aspects of software engineering in our research, and we know that to capture them, we need appropriate driving research questions and methods, as well as a focus on relevant stakeholders~\cite{Seaman99}. In this paper, we ask if and how we are making these provisions in our empirical studies. 

The focus of our investigation is how software engineering research approaches the inclusion and study of social aspects in software development. This led us to articulate questions about \textit{who} our research intends to benefit, \textit{what} are our research contributions, and \textit{how} methodologically we approach delivering relevant results given the \textit{who} and \textit{what} of our studies. 

We analyzed papers from two well-cited publishing venues to assess how much empirical software engineering research may explicitly consider or study social aspects.
We considered a cohort of papers (from 2017) published in the main technical track at the International Conference on Software Engineering (ICSE) and in the Empirical Software Engineering journal by Springer (EMSE). 
For these papers, we aimed to answer the following questions:

\begin{itemize}
\item \textbf{RQ1:}	Who are the \textbf{beneficiaries} (technical systems, human stakeholders, researchers) of the research contributions? 
Note that for papers that do not consider human stakeholders in their research goals, we would not expect the paper to directly study or address human and social aspects.
\item \textbf{RQ2:} What is the main type of \textbf{research contribution} (descriptive or solution) provided?
Descriptive papers add new or refute existing knowledge about a software engineering problem or context.
Solution papers present the design and/or evaluation of a new intervention (i.e., a process or tool). 
\item \textbf{RQ3:} Which \textbf{research strategies} are used? Some research strategies innately involve human subjects to collect or generate data, while others may rely on collecting previously archived, tool generated or simulated data.
\item \textbf{RQ4:} How do the reported research strategies \textbf{map} to the beneficiary and types of contributions in these papers? \end{itemize}

To answer our four research questions, we developed a socio-technical research framework to capture the main beneficiary of the research, the main type of research contribution, and the research strategies used. We refer to this framework as the Who-What-How framework.
We find that the majority of the 151 papers published in both venues in 2017 (ICSE and the EMSE journal) present research that the authors claim should \textbf{benefit human stakeholders} at some point in time, 
 but most of these do not use \textbf{research strategies} that directly involve human participants.
In terms of the \textbf{types of contributions}, we find that the majority of papers published at ICSE 2017 presented solutions (mostly technical) to address a software engineering problem, while the majority of papers published in the EMSE journal in 2017 are descriptive contributions that present insights about software engineering problems or how technical solutions are used. 
We conclude by calling for more diversification and triangulation of research strategies so that we may gain a deeper understanding of human and social 
aspects of software development practice to balance the current focus on the design and evaluation of technical innovations.
 
 The remainder of our paper is structured as follows. 
We discuss related work that has both informed and motivated this research in Section \ref{Background}.
In Section \ref{Framework}, we introduce the Who-What-How framework we designed for categorizing the beneficiaries, research contribution and research strategies in the papers we studied. 
In Section \ref{Methodology}, we present the methodology we followed to answer our research questions.  
In Section \ref{Findings}, we present the results from the four research questions we posed.   
We then interpret these results in Section \ref{Discussion}, discussing possible explanations and the implications of our findings.
We describe the limitations of our research in Section \ref{Limitations}. 
Finally, we conclude the paper by identifying areas for future work and list important takeaways in Section \ref{Conclusion}. 
Traceability artifacts from our analysis and a replication package are 
 at \replicationUrl.

\section{Background}
\label{Background}

The importance of social aspects in software engineering was recognized long ago~\cite{Weinberg1971, Brooks1975, Shneiderman1980, DeMarco1987}.
Typically there is at least one track on social aspects in the main research conferences, as well as special purpose workshops on the topic, such as the CHASE series\footnote{Cooperative and Human Aspects of Software Engineering, co-located with ICSE since 2008 \url{http://www.chaseresearch.org/}}. 
The papers presented at CHASE tend to address broad socio-technical topics, but the workshop focuses on early results. 
The Empirical Software Engineering and Measurement (ESEM) conference and the Empirical Software Engineering (EMSE) journal also attract papers that consider social aspects as their focus is on empirical methods, many of which directly involve human participants.
Some special journal issues have also addressed human aspects in software engineering using qualitative methods~\cite{DITTRICH2007, DYBAA2011}. 
Still, the debate about whether we study social aspects enough is ongoing and some researchers claim that coverage of these aspects is lacking~\cite{lenberg2014towards}. 

Beyond the discussion of \textit{how much} we study social aspects as part of software engineering research, is the discussion of \textit{how} we approach them methodologically. 
Researchers have focused on discussing specific methods (e.g., focus groups~\cite{kontio2008}, personal opinion surveys~\cite{Kitchenham2008}, or data collection techniques for field studies~\cite{Singer2008}) and providing guidelines on how to use them, or explaining the benefits and drawbacks of methods to assist in research design choices~\cite{Easterbrook2008}.
Seaman~\cite{Seaman99, Seaman2008} highlighted that a study focusing on social aspects asks different questions from one focusing on technical aspects, and needs to use appropriate methods that capture firsthand behaviors and information that might not be noticed otherwise. She discussed ways to incorporate qualitative methods into empirical studies in software engineering.

Social aspects can be approached methodologically by inferring behaviour from analyzing trace data of developers' past activities (e.g., code commits, code review comments, posted questions and answers on developer forums, etc.). But the analysis of trace data alone is fraught with threats to validity as it shows an incomplete picture of human behaviour, intent and social interactions in software engineering~\cite{Aranda2009, Kalliamvakou2014}. Furthermore, trace data alone cannot be used to predict how a new solution may perturb an existing process in industry settings~\cite{LMP16}, although relying on trace data can bring early insights about the feasibility of a solution design.
To appropriately capture and account for social aspects in software engineering research, we need to use dedicated methods that directly involve human participants in our empirical studies.

Method choice in empirical software engineering studies has been an item of reflection, especially around methods that are borrowed from other domains.
Zelkowitz~\cite{Zelkowitz2007} reported that researchers were using terms such as ``case study'' to refer to different levels of abstraction, making it hard to understand the communicated research.
However, as we are starting to recognize the misuse of certain methods, the research community is coming up with guidelines on how to use them correctly.  
Stol, Ralph, and Fitzgerald produced guidelines for grounded theory in the context of software engineering~\citeicse{Stol2016} as they found that many papers reporting the use of grounded theory lacked rigor. 
Runeson and H{\o}st adapted case study research guidelines to the software engineering domain \cite{Runeson2008}, also in part to address the misuse of the term ``case study'' in empirical software engineering studies. 
Sharp \emph{et al.}~\cite{Sharp2016} advocate using \textbf{ethnography} in software engineering studies as a way to capture rich insights about what developers and other stakeholders do in practice, \emph{why} they follow certain processes, or \emph{how} they use certain tools.

In this paper, we analyze published research in software engineering to investigate how social aspects are accounted for and studied.  
However, there have been previous efforts to reflect on and reexamine software engineering research to better understand where the community places value.  
For example, Shaw~\cite{Shaw2003} analyzed the content of both the accepted and rejected papers of ICSE 2002, as well as observed program committee conversations about which papers to accept. 
She found low submission and acceptance rates of papers that investigated ``categorization'' or ``exploration'' research questions, and low acceptance of papers where the research results presented were ``qualitative or descriptive models''.
A 2016 replication of Shaw's methodology \cite{Theisen2017} drew similar conclusions and identified a new category of papers, mining software repositories, was becoming common. 
While the earlier efforts focused on categorizing research and empirical studies in software engineering, the work we report in this paper focuses specifically on how studies approach social aspects and discusses the trade-offs and implications for the software engineering community's collective knowledge on the choices made in the studies we examined.
In the next section of this paper, we present a framework specifically designed for this purpose.

\section{A Socio-Technical Research Framework: The Who, What, How of Software Engineering Research}
\label{Framework}

\begin{figure}
    \includegraphics[width=\linewidth]{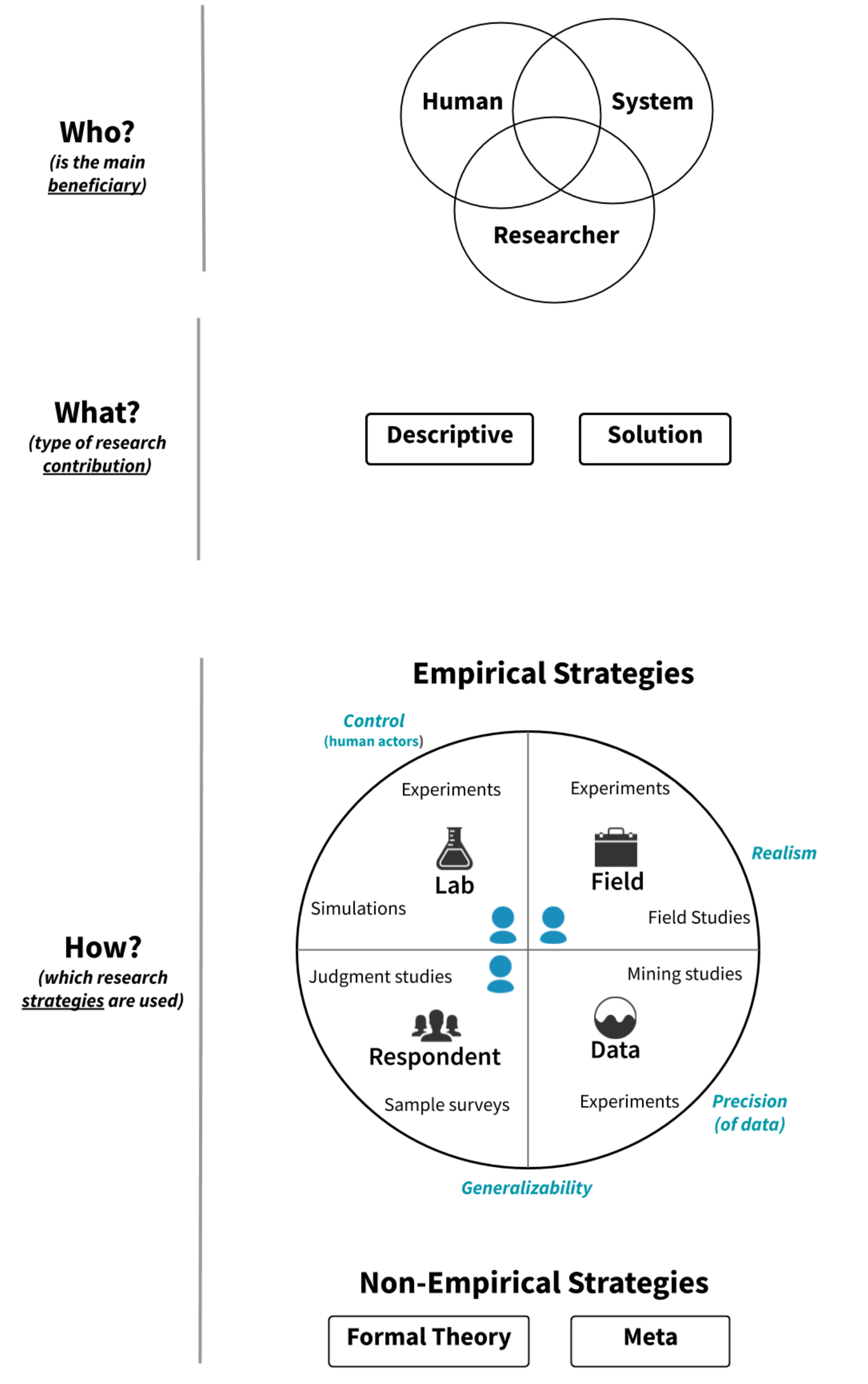}
    \caption{Our research strategy framework for categorizing the who (beneficiary), what (type of contribution) and how (research strategy) of empirical software engineering research. For the circumplex (circle) of empirical research strategies, note that in addition to labeling each quadrant with the type of strategy (Lab, Field, Respondent, and Data), we also show the main representative strategies in each quadrant (e.g., Experiment, Field Studies in the Field quadrant). We show \textbf{research quality criteria} (Control, Realism, Precision, and Generalizability) on the outside of the circumplex, positioned closer to the research strategies that have the highest \emph{potential} to increase those criteria in particular.}
    \label{fig:framework}
\end{figure}

To answer our research questions and guide our analysis of how empirical software engineering research papers may address human and social aspects in software engineering, we developed a socio-technical research framework to capture the main beneficiary of the research (the \emph{who}), the main type of research contribution in each paper (the \emph{what}), and the research strategies used (the \emph{how}).
The shape of our framework and the questions it poses emerged from several early iterations we followed when we tried to compare how papers address social and human aspects in software engineering research. 
Accordingly, our Who-What-How framework has three main parts, as shown in Fig.~\ref{fig:framework} and described below. 

\subsection{\emph{Who} Is the Research Beneficiary?}

The first part of the framework (see top of Fig.~\ref{fig:framework}) considers \emph{who} are the main beneficiaries of the research contributions claimed in a paper.  
All research papers intend for someone, or something, to be a primary recipient of the improvement or insight proposed by the research contribution, assuming that contribution is valid and practically relevant.
We consider the following possibilities for the beneficiary in our framework: 
\begin{itemize}
\item \textbf{Human stakeholders\footnote{We shorten this to ``Humans'' in the rest of the paper.}}, which may include software developers, architects, analysts, managers, end users and the social organizations they form; 
\item \textbf{Technical systems}, which may include tools, frameworks and platforms being used to support development\footnote{We recognize that most technical systems are studied or improved with the final goal to benefit a human stakeholder. However, we found in many papers that these human stakeholders are not discussed and that the research is aimed at understanding or improving the technical system.}; and
\item \textbf{Researchers}, which may include software engineering academics or industry research and internal tools teams. \end{itemize}

A paper's research contribution may be aimed at multiple beneficiaries.
For example, a paper may provide insights for researchers to consider in their future research studies, while at the same time make recommendations for practitioners to consider.
Likewise, a paper may provide insights that improve a technical system, such as improving the accuracy of a bug detection tool, but at the same time provide cognitive support to the developer who will use the tool.  
Alternatively, some papers may be clearly aimed at a single beneficiary. For example, we intend for the contributions of this paper to benefit researchers. 
\subsection{\emph{What} Is the Main Research Contribution Type?}

The second part of the framework (see middle of Fig.~\ref{fig:framework}) captures \emph{what} research contribution a paper claims.
To characterize the type of research contribution, we turn to a design science lens developed in previous work~\cite{Engstrom2019}.
Design science is a paradigm for conducting and communicating applied research, such as software engineering. Similar to other design sciences, much software engineering research aims to understand the nature of problems and real-world design contexts, that is \textbf{descriptive knowledge}, and/or  produce prescriptive knowledge to guide the design or improvement of \textbf{solutions} that address engineering problems.
Although some papers can have both a descriptive and a solution (prescriptive) contribution, we categorize papers according to a single main contribution, as emphasized by the authors of the paper.

\subsection{\emph{How} Is the Research Conducted?}

The last part of the framework (see bottom of Fig.~\ref{fig:framework}) helps us articulate \textit{how} the research was conducted by capturing the research strategies used. 
This part of the framework is derived in part from Runkel and McGrath's model of research strategies~\cite{runkel1972,mcgrath1995methodology}.
We describe the original Runkel and McGrath research strategy model and how we adapted it in the appendix of our paper (Appendix~\ref{AppendixA}).
A recent paper by Stol and Fitzgerald~\cite{Stol2018} also uses the Runkel and McGrath circumplex to provide consistent terminology for research strategies, which we also discuss in Appendix~\ref{AppendixA}.

The \emph{how} part of our framework first distinguishes empirical from non-empirical research.
Non-empirical research papers are not based directly on data (human or system generated)~\cite{gueheneuc2019empirical}.
Some non-empirical papers present literature reviews or meta-analyses of previous studies or empirically collected data, or they may describe research that uses formal methods, provides proofs or generates theories from existing theories. 
As we show later, the vast majority of papers published at ICSE and in the EMSE journal report or collect empirical data (some in addition to meta and/or formal theory research). This is not surprising as the EMSE journal in particular is aimed at empirical software engineering research.

There are four empirical strategy quadrants that we show embedded in a circumplex in Fig.~\ref{fig:framework}. 
In addition to labeling each quadrant with the type of strategy (Lab, Field, Respondent, and Data), we also show two representative strategies for each quadrant (e.g., Experiment and Field Studies in the Field quadrant). 
It is important to not confuse research strategy with research method.
A research method is a technique used to collect data. For example, interviews may be used as a method for collecting information from field actors as part of a field study, or as a method for collecting data as part of a sample survey~\cite{Easterbrook2008}. 
In contrast, a research strategy is a broader term~\cite{Stol2015} that may involve the use of one or more methods for collecting data. It indicates how data is generated, whether directly or indirectly from participants or produced by the researcher, and suggests the setting used for the study.
    
Three of the empirical strategy quadrants (lab, field and respondent strategies) directly involve human participants (represented by a person icon in the framework). 
While strategies that fall in these three quadrants all involve human participants, they vary significantly in terms of the study setting. 
Field studies occur in the context of work. For example, a software company, or a classroom when studying the educational aspects of software engineering. 
Lab studies are conducted in a contrived context, often in a university or research center, 
and respondent studies are conducted in settings of convenience, such as a workplace, home, classroom or conference.

The fourth empirical research strategy quadrant, which we refer to as \emph{data}, captures studies that are conducted \emph{in silico}\footnote{By \emph{in silico}, we mean performed on a computer or via computer simulation.} and do not directly involve human participants, although they may use previously generated human or system data that is behavioral.

Every empirical research strategy has strengths and weaknesses to consider in terms of \textbf{research quality criteria}~\cite{runkel1972}, and every study design decision leads to trade-offs in terms of the potential to increase or decrease quality criteria.
We consider four quality criteria: 
\begin{itemize}
\item \textbf{Generalizability} of the evidence over the population of human or system actors studied;
\item \textbf{Realism} of the context where the evidence was collected and needs to apply; 
\item \textbf{Control} of extraneous human behaviour variables that may impact the evidence being collected; and
\item \textbf{Precision} of the system data that is collected as evidence.
\end{itemize}
The four criteria are shown on the outside of our empirical strategy circumplex in Fig.~\ref{fig:framework}, and they are positioned in proximity to the quadrants whose strategies have the potential to increase those criteria (but doing so is not a given, as we describe below). 
Since all strategies have inherent strengths and weaknesses, it may be important to \textbf{triangulate} across research strategies to mitigate the weaknesses of using a single strategy~\cite{runkel1972} in one's research. 

Next, we describe research strategies that belong to the four different empirical quadrants in more detail. We discuss how the strategies show potential for higher or lower realization of the research quality criteria listed above.

\subsubsection{Field Strategies}
Field strategies involve researchers entering a natural software development setting to study the socio-technical system in action. This includes the technical systems used to support engineering activities, the system or project under development, and the human stakeholders, such as developers, engineers and managers.

A \emph{field study} is where the researcher observes (e.g., through an ethnography) study subjects without any explicit interventions in place.
With a field study, realism is high but control over human activities is low, and generalizability is low as only a limited number of companies are typically considered.
For example, a researcher may observe how agile practices are adopted in a startup company, leading to descriptive insights about the practice in a realistic but specific setting.

A \emph{field experiment} occurs in a natural development setting but one where the researcher controls certain aspects of the setting and may aim to compare the impacts of different solutions or environment conditions.  
A field experiment may be more obtrusive than a field study, and thus will lower realism, but it has the potential for higher control over human participants' activities.
For example, a field experiment may involve the comparison of 
a novel automatic testing tool to an existing tool in the developers' realistic and natural development setting. 
A study that only considers data traces from the field (e.g., from a data mining study or use of machine learning in an experiment) is categorized as a data study (see below) as these studies do not involve the direct involvement or observation of human participants in their natural environment.

\subsubsection{Lab Strategies}
 Lab strategies typically involve testing hypotheses in highly controlled situations or contrived environments with human participants and technical systems as actors.  
Control of human actor activities may be achieved but at the expense of realism (the setting is contrived), and it may be more difficult to achieve generalizable results.

A \emph{lab experiment} is one lab strategy where the experimenter controls the environment and interventions or tools used. For example, a researcher may investigate the effects of a new debugging tool in comparison to the \emph{status quo} debugger on programming task efficiency using graduate students as participants in a lab setting.

In comparison, an \emph{experimental simulation} may try to increase realism by setting up an environment that mimics a real environment in industry.
For example, a researcher may investigate different modalities for project management meetings in an environment that is similar to a collaborative meeting room in a company.
Doing so increases realism but at the expense of control over variables that may come into play in a simulated environment where human actors may have more freedom to act naturally. 
Of note is that many lab strategy studies tend to use students in software engineering research. As part of our analysis, we also noted which of these studies involved practitioners and mention this result when we discuss the implications of our findings.

\subsubsection{Respondent Strategies}
Respondent strategies are often used in software engineering research to gather insights from practitioners and other stakeholders. 
The ability to collect data from many participants using respondent strategies has the potential for higher generalizability (if a broad and large sample is recruited), but at the expense of lower realism as factors that may influence each individual participant's responses cannot be observed or anticipated.

A \emph{sample survey} is a respondent strategy which may, for example, involve an online questionnaire or a set of interviews to gather opinions from human participants.  
For example, a questionnaire or interviews may be used to learn how developers perceive code quality, or to learn how continuous integration tools are used by developers and which challenges they encounter using these tools.

A \emph{judgment study}, another respondent strategy, asks human participants to give their expert opinion on the effectiveness of a new tool or process. Typically participants try out a new tool or process in a setting of convenience.  
For example, a researcher may wish to ask developers for their opinion about a new debugging tool by asking them to use it briefly in place of their regular tool and to provide an opinion.
Judgment studies are distinguished from laboratory and field strategies by the use of an environment of convenience and by the lower control in how the tool is used. 

\subsubsection{Data Strategies}
Data strategies refer to empirical studies that rely primarily on archival, generated or simulated data.
The data used may have been collected from a naturally running socio-technical system, or it may be partially generated in an experiment.
This type of research strategy is frequently used in software engineering research due to the technical components in software engineering, as well as the widespread availability of extensive trace data and operational data from modern software development tools.

Data strategy papers may use a wide range of specific methods, including \emph{experiments}, to evaluate and compare software tools (e.g., such as a defect prediction tool) with historical data sources.
Data strategy papers may also refer to \emph{data mining studies} used to gather descriptive insights about a socio-technical system. 
They are sometimes used to infer human behaviours from human-generated, behavioral trace and operational data (e.g., source code, commit comments, bugs). However, as other researchers have reported, this data is inadequate for understanding previous human behaviour~\cite{Aranda2009} and can be misleading in terms of predicting future developer behaviours~\cite{LMP16}.
Data strategy papers have lower control over human behavioral variables but have the potential for higher precision of system data. 
They may have potential for higher realism (if the data was collected from field sites) and higher generalizability (if the project data is from many projects).

In the next two sections of the paper, we describe how we used the Who-What-How research framework to answer our research questions.  
Later, we discuss how the framework may be used to reflect on and guide software engineering research.

\section{Methodology}
\label{Methodology}

To answer our research questions (see Section \ref{cha:introduction}), we read and analyzed ICSE Technical Track and EMSE journal papers from 2017. 

We selected the ICSE venue for our analysis because it is considered the flagship conference in software engineering, and because one would expect it represents the broad areas of software engineering research. We considered EMSE as it is one of the top two journals in software engineering and focuses on empirical studies. We selected these venues because they both cover a broad set of topics---we selected EMSE in particular because of its focus on empirical software engineering research. We do not aim for broad coverage of all SE venues, but rather two exemplar flagship venues that can illustrate our framework.

ICSE 2017 had 68 papers in the technical track, and EMSE published 83 papers in 2017 (including special issues but excluding editorials). This resulted in a dataset of 151 papers.

As we read the papers, we answered the questions posed using the framework described above in Section~\ref{Framework} and recorded our answers in a shared online spreadsheet (available in our replication package at \replicationUrl).
This process was highly iterative and our framework (and coding categories) emerged from earlier rounds.
In particular, the need to consider ``precision''  as a quality criterion in our framework emerged when we realized that many of the papers published in software engineering rely on datasets procured from technical components that offer precise measurement. 
When the framework had stabilized to its current form, each paper was read and coded by at least two (but often three or more) members of our research team.
Any differences in our responses to the framework questions (e.g., should this paper be described as benefiting researchers, systems, or humans) were discussed to decide what the most appropriate answer should be. We did this by reading the paper again, and for some, recruiting an additional member of our research team to read and discuss the paper.  
Initial disagreements were easy to resolve as additional information was often contained in an unexpected section of a paper (e.g., a small judgment study may have been done but not discussed until later in the paper). The process was laborious as sometimes research strategies or beneficiaries were mentioned in unexpected places.
We always relied on the paper author's own words to justify our choices. For example, a paper had to explicitly refer to a human beneficiary (``developers'' or ``programmers'', for instance) for us to justify coding it as such.
Our selected answers, with quotes or comments to justify coding that was more subjective or saw some disagreement, can be found in our artifact package (see \replicationUrl).

\textbf{RQ1} is to identify \emph{who} is the main beneficiary of the research. To answer \textbf{RQ1} we considered both the framing of the research questions and the introductions of the papers. In some cases, it was difficult to answer this question and we had to refer to the discussion and/or conclusion of the papers to identify who or what was the claimed or intended beneficiary of the research.
In the case where we identified that a human stakeholder was a stated beneficiary, we copied a quote from the paper into our analysis spreadsheet.
We deemed something as involving human beneficiaries (as `human' or `both') if the paper contained a statement referring to human stakeholders. To find this mention, we read each paper, augmenting our reading with keyword searches for `developer', `human', `user', `tester', `engineer', `coder', and `programmer'. 

Mention of humans in the paper did not always include an in-depth discussion of how a human might benefit from the tool or process. Some papers only tangentially mentioned that humans might benefit from their approach. 
For example, Lin \emph{et al.}'s~\citeicse{Lin2017} ICSE 2017 data strategy paper, ``Feedback-Based Debugging'', mentions that their ``approach allows developers to provide feedback on execution steps to localize the fault''. 
By comparison, a paper we coded as referring to a system as the only stakeholder noted that ``LibD can better handle multi-package third-party libraries in the presence of name-based obfuscation'' (Li \emph{et al.}~\citeicse{Li2017}). 

To answer \textbf{RQ2} and to identify \emph{what} the main research contribution is, we read the abstract, introduction, results/findings, discussion and conclusions of each paper. 
Again, categorizing a paper as a solution or descriptive paper was not always straightforward, but we relied on how the authors framed their results to decide which was the main contribution.
When a descriptive paper concluded with a proposed solution that was not evaluated, we coded it as a descriptive paper---in fact, the solution was often provided as a way to justify the impact of the descriptive results reported.
Many solution papers had a (usually) small descriptive contribution, thus coding as both descriptive and solution would not allow us to discriminate the contributions in the papers we analyzed.
 For example, a paper that provides a theory of how continuous integration improves software quality or describes challenges with using continuous integration tools would be categorized as descriptive, whereas a paper that proposes and/or evaluates a new continuous integration tool would be categorized as a solution paper. 
 
To answer \textbf{RQ3} and to identity \emph{how} the research was conducted in terms of the research strategy used, we focused on the methodology sections of the papers. 
Again, we used our framework to help distinguish different kinds of strategies.
For each paper, we noted if one or more strategies were used.
Sometimes an additional strategy was mentioned as a small additional step in the reported research in the discussion or background sections of the paper, so we also read the entire paper to be sure we captured all the strategies reported.
Although not one of our main research questions, we also coded which papers directly involved industry practitioners in their studies. 
We discuss this finding in the discussion section of this paper.

To answer \textbf{RQ4} and to identify how the reported research strategies \emph{map} to the beneficiary and type of research contribution, we mapped our responses in our spreadsheet and visualized the results.
For this question, we expected to see that papers with a human beneficiary may be more likely to also use research strategies that directly involve and control for human behaviours.
In terms of how research contributions map to beneficiary and research strategy, we were curious to see if there were some patterns in this regard as we did not have an initial expectation about this mapping.

For the purposes of replication and traceability, we provide our methodological tools, the anonymized raw data, and analysis documents and spreadsheet at \replicationUrl. 

\section{Findings}
\label{Findings}
We present the findings from applying the Who-What-How framework to ICSE conference and EMSE journal papers published in 2017.  
We interpret and discuss the possible implications of these findings later, in Section \ref{Discussion}.

\subsection{RQ1: Who are the \textbf{intended beneficiaries} of the published research?} 
\label{RQ1}
\begin{figure}[b]
    \centering
    \includegraphics[width=\linewidth]{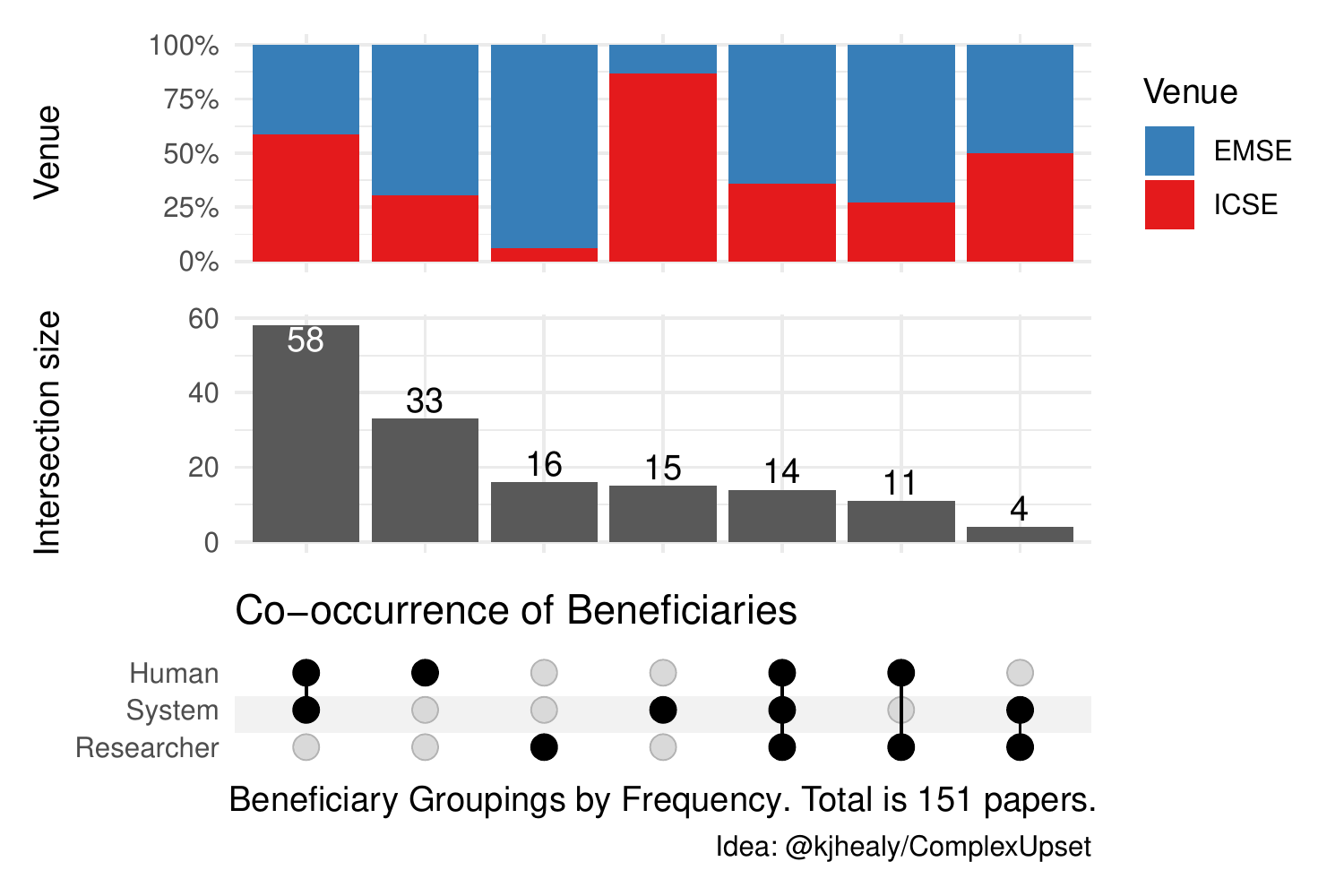}
    \caption{Intended beneficiaries of the research contributions. This upset plot \cite{Lex2014} captures how our beneficiaries overlap. The bottom filled/unfilled circles represent set membership, i.e., a filled circle indicates papers we coded with that beneficiary. Proceeding vertically up the diagram, the intersection size represents the overall number of papers with those memberships. For example, papers with \textsf{Human\&System} beneficiary were 58/151 of the papers in the data we coded. The Venue plot shows what proportion of those papers were found at EMSE, or at ICSE. For the \textsf{Human\&System} papers, over half were published at ICSE. On the other hand, we see EMSE in 2017 published more papers with only \textsf{Researcher}s, while ICSE 2017 published more papers with \textsf{System} as the only beneficiary. }
    \label{fig:ben_venn}
\end{figure}

Figure \ref{fig:ben_venn} shows an upset plot \cite{Lex2014} of the intended beneficiaries. 
We were liberal in coding the possible beneficiaries. 
For a paper to be coded as \textsf{Human}, we checked if an author noted their paper described a tool that could solve a \emph{human} problem (e.g., improve developer productivity) or benefit human or organizational stakeholders. For example, in the 2017 EMSE journal paper, ``A robust multi-objective approach to balance severity and importance of refactoring opportunities''~\citeicse{Mkaouer2016}, the primary contribution is for a \textsf{System} (Note: all emphasis is ours): \begin{quote}``The results provide evidence to support the claim that our proposal enables the \emph{generation of robust refactoring solutions} without a high loss of quality using a variety of real-world scenarios."\end{quote} However, we also identified a claim for human benefits in this paper: \begin{quote}``The importance and severity of code fragments can be different after new commits introduced by \emph{developers}. [...] the definition of severity and importance is very subjective and depends on the developers’ perception." \end{quote}

Likewise, if a paper that predominantly studied human behaviours mentioned their findings may improve a tool, we coded the paper as also benefiting a \emph{system} component. We had a flexible interpretation of `tool'. For example, in the ICSE 2017 paper, ``How Good is a Security Policy against Real Breaches? A HIPAA Case Study"~\citeicse{Kafali2017}, the tool in question is ``a formal representation of security policies and breaches [and] a semantic similarity metric for the pairwise comparison of norms", but the beneficiaries include human stakeholders: \begin{quote}``Our research goal is to help \emph{analysts} measure the gaps between security policies and reported breaches by developing a systematic process based on semantic reasoning."\end{quote}

One could conclude that all papers benefit researchers in some way. However, we coded papers as benefiting \textsf{Researcher}s when a main focus of a paper contribution was clearly aimed at researchers
(e.g., in the case of a benchmark for future research use). 
We found that EMSE reports more systematic literature reviews, papers that lead to artifacts, and benchmarks aimed at researchers than ICSE does.

Since our coding permitted one or more of our three beneficiary types (Human, System, Researcher), we can clearly see multiple beneficiaries in the upset plot---\textsf{for All, Human\&System, Human\&Researcher}, and \textsf{System\&Researcher}.
We found that the majority of papers from both venues claim human stakeholders as a possible beneficiary. 
Specifically, 77\%/76\% of the EMSE/ICSE papers (respectively) claim their research may benefit human stakeholders---many also claim technical systems and/or researchers as beneficiaries---but we find that more of the EMSE papers (27\%) claim humans as the sole beneficiary compared to 15\% of ICSE papers.

\subsection{RQ2: What type of \textbf{research contributions} are provided?} 
\label{RQ2}
Table~\ref{tbl:contrib} illustrates how many \textsf{Descriptive} vs \textsf{Solution} papers we captured using the framework. 
Note that many of the papers we coded as solution papers also had a descriptive contribution, either in terms of problem understanding or solution evaluation. For example, ``Code Defenders: Crowdsourcing Effective Tests and Subtle Mutants with a Mutation Testing Game" from ICSE 2017~\citeicse{Rojas2017} describes problems with generating effective mutation tests. We coded this as \textsf{Solution} since the main contribution is a code defender multiplayer game for crowdsourcing test cases. However, in the process of conducting the empirical study, the paper describes many of the problems with the approach.

If the solution contribution was minor (e.g., a recommendation for a new tool following a mostly descriptive paper), we coded the paper as a \textsf{Descriptive} paper. For example, we coded ``An initial analysis of software engineers' attitudes towards organizational change''~\citeicse{Lenberg2016} from the 2017 EMSE journal as a \textsf{Descriptive} study using a respondent strategy based on a survey to describe attitudes. One outcome of the paper, however, is: \begin{quote}``[a] proposed model [that] prescribes practical directions for software engineering organizations to adopt in improving employees’ responses to change".\end{quote}

Across both venues, 43\% of papers were \textsf{Descriptive}, and 57\% presented \textsf{Solutions} (see Table~\ref{tbl:contrib}).
More ICSE papers were identified as \textsf{Solution} papers, and most solutions were technical in nature. 
ICSE and EMSE published 81\%/37\% \textsf{Solution} papers and 19\%/63\% \textsf{Descriptive} papers, respectively.  
This large difference in contribution type mirrors results by Shaw \emph{et al.}'s analysis of ICSE 2002 papers~\cite{Shaw2003} and the replication of her study in 2016~\cite{Theisen2017} that found lower submission and acceptance of papers with results that were seen as ``qualitative and descriptive models''. 

\begin{table}[h]
\centering
\caption{Counts and Proportions of Research Contributions Per Venue}
\label{tbl:contrib}
\begin{tabular}{lcccccc} 
\toprule
Purpose     & \multicolumn{2}{c}{All} & \multicolumn{2}{c}{ICSE} & \multicolumn{2}{c}{EMSE}  \\
            & Count & Proportion      & Count & Proportion       & Count & Proportion        \\ \midrule
Descriptive & 65    & 0.43            & 13    & 0.19             & 52    & 0.63              \\
Solution    & 86    & 0.57            & 55    & 0.81             & 31    & 0.37              \\
\bottomrule
\end{tabular}
\end{table}

\subsection{RQ3: Which \textbf{research strategies} are used?} \label{RQ3}

\begin{figure}[htb]
    \includegraphics[width=\linewidth]{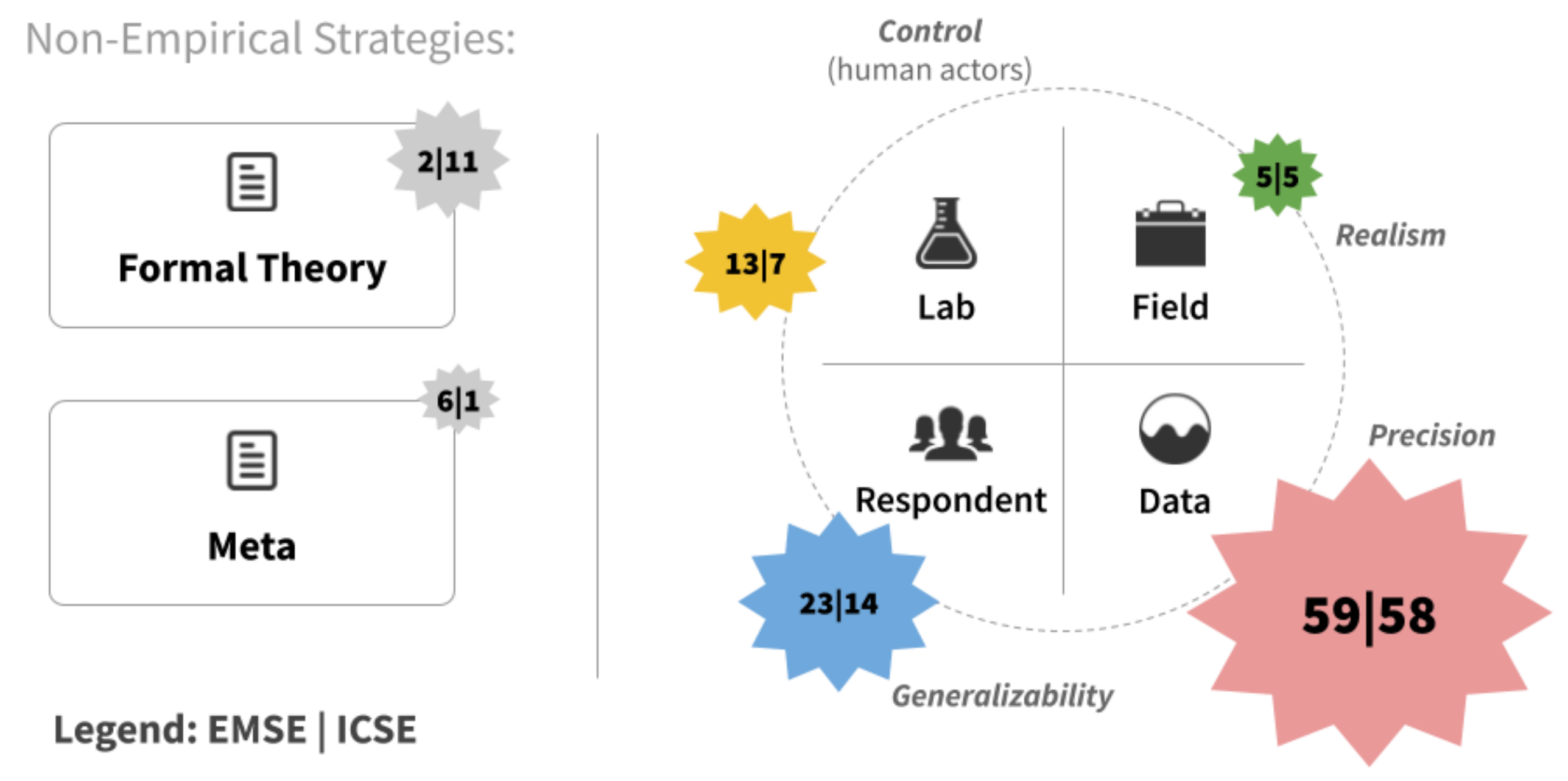}
    \caption{Counts of the research strategies used in the EMSE/ICSE 2017 papers, respectively. Note: Some papers reported more than one strategy.}
    \label{fig:smsresearchstrategies}
\end{figure}

Figure \ref{fig:smsresearchstrategies} shows the framework quadrants for research strategies. We report totals for EMSE and ICSE separated by a vertical pipe character. 
For example, we identified 23 EMSE and 14 ICSE papers using respondent strategies (shown on the figure as 23$|$14). 
Since some papers report two strategies, the totals add up to more than the 151 papers we analyzed. Section \ref{sec:triangulate} expands on how some papers use triangulation of research strategies.

Among the 151 papers we examined, we found a higher use of \textsf{Data} strategies (59 EMSE $\vert$ 58 ICSE) compared to any of the other research strategies (see Fig. \ref{fig:smsresearchstrategies}).

We expand on how we classified papers with specific examples. We provide the complete classification in our replication package available at \replicationUrl{}. A short sample is provided in Appendix \ref{Appendix B}.

 For \textsf{Data} strategies, an example is from Christakis \emph{et al.}'s ICSE 2017 paper, ``A General Framework for Dynamic Stub Injection''~\citeicse{Christakis2017}. This paper describes their novel stub injection tool and subsequent evaluation by running it on a series of industry applications which they used to instrument system calls and monitor faults. Since this paper generates data on system faults based on their tool, we classified this as a \textsf{Data} strategy.
 
A second example of a paper we coded as \textsf{Data} strategy is Joblin \emph{et al.}'s ICSE 2017 paper, ``Classifying Developers into Core and Peripheral: An Empirical Study on Count and Network Metrics''~\citeicse{Joblin2017}, which analyzes commit data from GitHub projects to study aspects of human behavior and uses prediction algorithms to classify developers as core or peripheral in open source GitHub projects. This descriptive study relies on GitHub trace data. 

There were significantly fewer instances of the other empirical research strategies: \textsf{Field} (5 EMSE $\vert$ 5 ICSE), \textsf{Lab} (13 EMSE $\vert$ 7 ICSE), and \textsf{Respondent} (23 EMSE $\vert$ 14 ICSE). We discuss the possible implications of this imbalance of research strategy use in Section~\ref{Discussion}, but first we give examples of each of the non-data quadrants shown in Fig.~\ref{fig:smsresearchstrategies}. We also report the number of papers we identified for each strategy (totals and EMSE$|$ICSE). 

\subsubsection*{Field Strategies} 

We identified 8 papers ($4|4$) that conducted \textsf{field studies} (total of both venues). Heikkilä \emph{et al.}~\citeicse{Heikkil2017} conducted a large field study at Ericsson examining requirements flows for a paper in the 2017 EMSE journal. Since the interviews were done with Ericsson developers in their natural work environment, realism was potentially high for this field study: \begin{quote}``We present an in-depth study of an Ericsson telecommunications node development organization .... Data was collected by 43 interviews, which were analyzed qualitatively.''\end{quote}

\textsf{Field experiments} were relatively uncommon (2 in total, 1$|$1). In their ICSE 2017 paper, He \emph{et al.}~\citeicse{Jiang2017} report a data study followed by a field experiment when studying test alarms. They integrated their tool into the system of their industry partner and observed the results of practitioners using the tool. This significantly increased the realism of the study.

\subsubsection*{Lab Strategies} 

For \textsf{laboratory experiments}, we identified 16 papers in total (11$|$5).
An example is Charpentier \emph{et al.}'s 2017 EMSE journal paper, ``Raters' reliability in clone benchmarks construction"~\citeicse{Charpentier2016}. The study used a combination of experts and students in a lab setting to rate software clone candidates, which were then used to evaluate clone detection tools. While there was a portion of the strategy in which respondents---their subjects---were asked to evaluate the clones, this is not a respondent strategy but rather a particular method used in the lab experiment. The experimenters controlled the conditions under which clones were evaluated in order to evaluate the independent variable of rater experience.

\textsf{Experimental simulations} occurred 4 times ($2|2$). In their ICSE 2017 paper, ``Do Developers Read Compiler Error Messages?''~\citeicse{Barik2017}, Barik \emph{et al.} conducted an eye-tracking study of students at their institution. Participants were asked to use error messages to debug software programs, and eye-tracking hardware was used to understand how participants solved the problem. Since this was not a controlled experiment, we coded this as an experimental simulation (a Lab quadrant strategy) since the main objective was to simulate, to some extent, the {realism} of actual debugging.

\subsubsection*{Respondent Strategies} 

 Online questionnaires (surveys) and interviews were common ways to implement a \textsf{sample survey} strategy in our sample. Sample surveys occurred 21 times (16 EMSE $|$ 5 ICSE).

The NAPIRE survey series, published in the 2017 EMSE journal and authored by M{\'{e}}ndez \emph{et al.}~\citeicse{Fernndez2016}, is a good example of using a series of online questionnaires to understand the problems practitioners have with requirements engineering. The surveys are broadly distributed to maximize generalizability from respondents.

Hoda and Noble's ICSE 2017 paper, ``Becoming Agile: A Grounded Theory of Agile Transitions in Practice''~\citeicse{Hoda2017}, used interviews to gather responses from various developers about their experiences transitioning to Agile development methods in their work. Because these researchers gathered responses from a wide variety of developers working in different settings, this increased the generalizability of their findings to other development contexts beyond those represented in the studies. Had they instead focused on a single company in detail, we would have coded this as a field study rather than a respondent strategy (sample survey).

The other type of respondent strategy is the \textsf{judgment study}, which occurred 16 times ($7|9$). This strategy can be seen in Bezemer \emph{et al.}'s 2017 EMSE journal study, ``An empirical study of unspecified dependencies in make-based build systems"~\citeicse{Bezemer2017}. In this paper, the authors began with a data strategy and then asked for professional feedback on the results: \begin{quote}``We contacted a GLIB developer to validate the patches that we submitted."\end{quote} Similar studies might post results as pull requests (for bug fixes, for example) and monitor acceptance rates.
Many of the judgment studies we saw involved few participants so these studies did not realize the high potential for generalizability that could have been achieved with a respondent strategy. 

\subsubsection*{Non-Empirical Strategies} 

For the non-empirical strategies, 13 papers ($2|11$) contained an aspect of formal methods or developed a \textsf{formal theory}, such as Faitelson and Tyszberowicz's ICSE 2017 paper, ``UML Diagram Refinement (Focusing on Class- and Use Case Diagrams)''~\citeicse{Faitelson2017}.
A total of 7 non-empirical strategy papers ($6|1$) were \textsf{meta} papers, i.e.,  aimed at other researchers
(such as systematic literature reviews or discussion of research methods). For example, ``Robust Statistical Methods for Empirical Software Engineering'' by Kitchenham \emph{et al.} in the 2017 EMSE journal~\citeicse{Kitchenham2016} aims to: \begin{quote}``explain the new results in the area of robust analysis methods and to provide a large-scale work example of the new methods.''\end{quote}

\begin{figure}[]
\begin{subfigure}{\textwidth}
    \centering
    \includegraphics[height=3.2in]{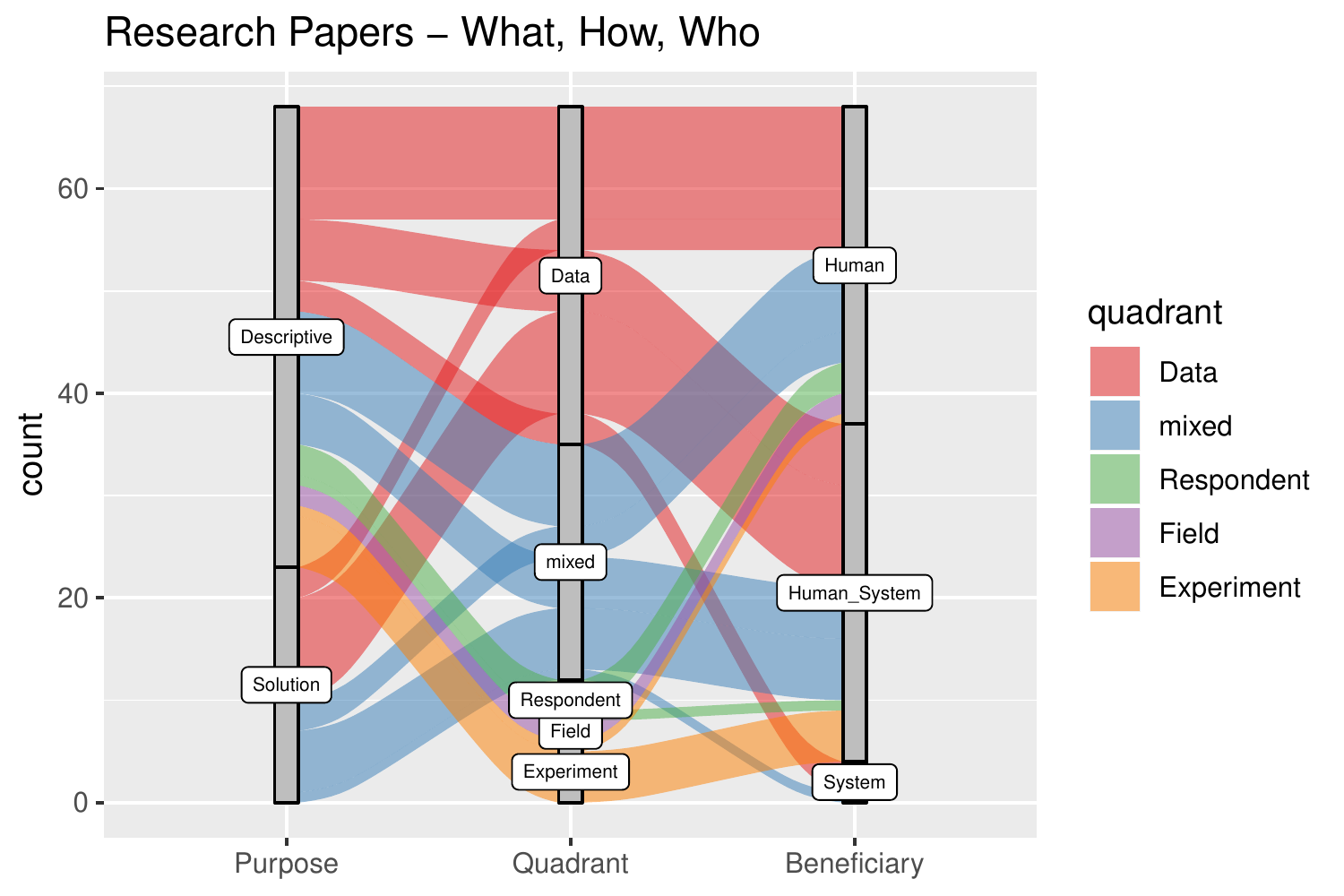}
      \caption{EMSE alluvial diagram}
		\label{fig:sankey:emse}
 	\end{subfigure}
 \begin{subfigure}{\textwidth}
    \centering
    \includegraphics[height=3.2in]{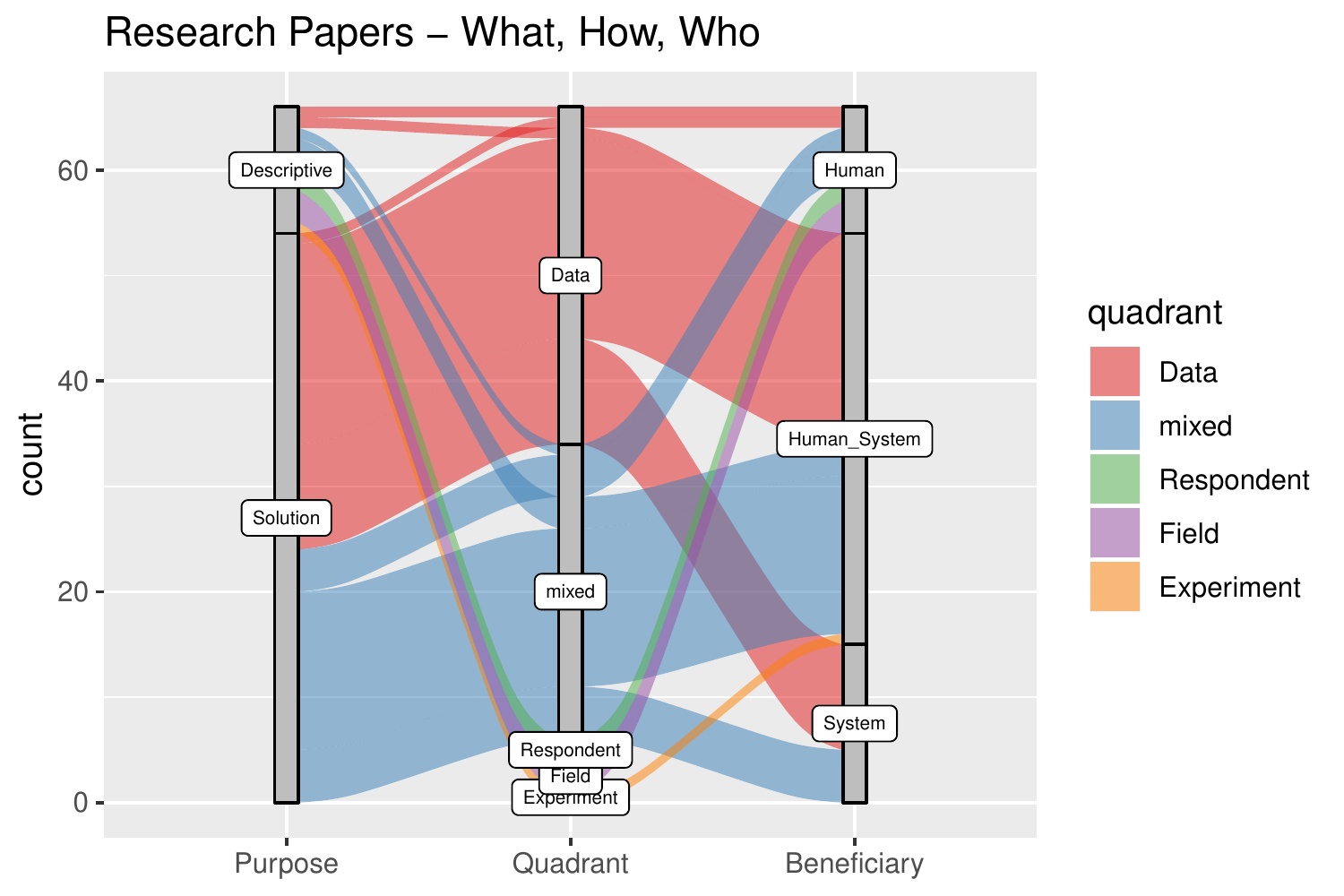}
      \caption{ICSE alluvial diagram}
\label{fig:sankey:icse}
    \end{subfigure}
    \caption{These two alluvial diagrams capture flows from human and system beneficiary (Who), to empirical research strategy quadrant (How), to purpose (What). The size of the lobes in each strata (column) reflects proportion of classified papers (for example, there were few papers classified as benefiting researchers, in the left-most column). The width of the alluvia (flow lines) likewise captures proportion and is colored by the chosen research strategy captured in the legend. For the \textbf{EMSE papers} (top, Fig. \ref{fig:sankey:emse}), we see an emphasis on human beneficiaries but more use of data strategies producing more descriptive contributions. For the \textbf{ICSE} papers (bottom, Fig. \ref{fig:sankey:icse}) we see a more even emphasis on human and system beneficiaries but more use of data strategies over other strategies, leading to more solution oriented contributions.}
 	    \label{fig:sankey}
\end{figure}

\subsection{RQ4: How do the reported research strategies \textbf{map} to the beneficiary and types of contributions in the 2017 EMSE and ICSE papers? } 
\label{RQ4}

To illustrate the mappings from research strategy to research contribution type and beneficiary, we created the alluvial flow diagrams shown in Fig. \ref{fig:sankey}. 
These diagrams outline the mappings between research beneficiary (RQ1), research quadrant (RQ3), and research purpose (RQ2) for the two venues. 
We show only the human and system beneficiary categories on the left side, omitting the few papers that focus on researchers only as beneficiary (to improve the clarity of our diagram).

The alluvial diagrams 
show that both venues report a high use of data strategies (the how) despite some key differences in beneficiaries (who) and contribution types (what).
That said, there is more use of non-data strategies (that is, strategies that directly involve human participants) in the EMSE papers (47\% of EMSE empirical papers, compared to 36\% of the ICSE empirical papers). 
This may be because more EMSE papers aimed at human stakeholders as the sole claimed beneficiary. 

For ICSE papers, the majority of non-data strategy studies map to descriptive research contributions, and the majority of data strategy studies map to solution contributions. 
For the EMSE papers, by contrast, we see that many of the data studies map to descriptive contributions (EMSE has fewer solution papers), but we note that many of these descriptive contributions aimed at humans as the beneficiary.
We discuss the possible implications of not involving human actors in studies that are aimed at human beneficiaries later in the paper (see Section~\ref{downsides}).

\section{Interpreting Our Findings}
\label{Discussion}

This section presents our interpretation of the findings, with key insights highlighted in bold.
First,  
we discuss why we see an emphasis on data strategies in software engineering research (for both venues), and how data strategies may be limited for understanding and experimenting with the human and social aspects in software engineering. 
Then, we discuss how research strategies are not triangulated as much as we may expect in papers presented in these venues, and how more triangulation of research strategies could bring more attention to 
human and social aspects in 
software engineering research and practice.

\subsection{A Penchant for Data Strategies}

We found that the majority of the ICSE conference and EMSE journal papers we analyzed relied on data strategies (71\% and 85\%, respectively), and over half of the total papers 
 relied solely on data strategies in their choice of method.
There are several reasons that may help explain why data strategies are commonly used in software engineering research.

Data strategies are \textbf{well-suited to understanding and evaluating technical aspects} for papers that focus on technical systems as the beneficiary. Many authors in the set of papers we analyzed aimed to improve a technical system or component as one of their main research beneficiaries. 
We found that most of the ICSE papers we analyzed are solution papers, and data strategies are used in many solution papers to show the feasibility and scalability of technical solutions.  
 And for descriptive papers---which the majority of the EMSE papers we analyzed can be considered---much can be learned from data about technical systems, and valuable knowledge of human and social aspects can also be gleaned from data alone.

In recent years, there is \textbf{increased availability of data} from software repositories and diverse data sources concerning software projects.  
These data sources encompass a rich resource concerning both technical and human aspects for conducting empirical software engineering research.
We have access to open or proprietary source code when research collaborations are in place; development data, such as issues, bugs, test results, and commits; crowdsourced community knowledge repositories, such as Stack Overflow and Hacker News; and operational and telemetry data from the field, such as feature usage, A/B testing results, logging data, and user feedback~\cite{Hassan2008}.
Note that the analysis of ICSE papers from 2016 by Thiesen \emph{et al.}~\cite{Theisen2017} also showed an increase in mining software repository papers (one type of data strategy paper) over Shaw's earlier study on ICSE 2002 papers~\cite{Shaw2003}.

\textbf{Data analysis is a core skill} that many computer scientists and software engineering researchers possess as they are more likely to come from scientific, engineering and mathematical backgrounds, with expertise in statistics, data mining, natural language processing and AI.
In addition, the emergence of new and powerful machine learning and AI techniques that scale to software engineering projects and is taught as part of software engineering curricula also may help explain why data strategies are used so frequently in our field.

Data strategy studies have the potential to achieve high \textbf{generalizability}, particularly when multiple projects are studied.
The analysis of data from repositories hosted on sites such as GitHub may be more generalizable to a broader set of projects.
Similarly, using data from real-world projects has the potential to increase \textbf{realism}, particularly in terms of the technical systems studied. 

Data strategies also lend themselves naturally to \textbf{replication}, a desirable aspect to improve scientific rigor. 
We are starting to see more data strategy papers in software engineering that are accompanied (either as a requirement or optionally) by replication packages that include the software artifact data, algorithms, scripts, and other tools used in the studies. 
These packages are recognized by the community as having high potential for replication. 
Some evidence further suggests that replication packages and open science leads to more citations~\cite{Colavizza:2019aa}, which might also motivate this strategy choice.
Other strategies (such as lab, field, and respondent) can also provide replication packages, but exact replications are more difficult when human participants are involved as their behaviour can be impacted by more nuanced variables, and their behaviour is not as predictable as technical components in studied socio-technical systems.

As our answer to RQ4 shows, many data papers aim at humans as a beneficiary, and while some triangulate using another strategy that involves human participants, many do not directly study humans at all. We discuss the implications of this below, but first discuss why an over reliance on data strategies may be a problem.

\subsection{The Downside of Using Data Strategies for Studying Human Aspects}
\label{downsides}

Our analysis found that many papers in both venues we inspected claim their research would benefit humans in some way, but did not directly involve them in their studies: 43\% and 54\% for EMSE and ICSE (respectively) claim they benefit humans but did not directly study them, instead relying on data traces.

In some cases, this may be justified. 
For example, a study that evaluates a new technique to improve the build time of a project may not benefit from human feedback as the faster compilation probably implies a better developer experience.

But many proposed {solutions} may need to be evaluated in a human stakeholder's context.
For example, a solution paper that proposes a recommender system of possible defects may need to be evaluated with human actors to see how the tool perturbs the context in which it is used~\cite{LMP16}, and to {control} for other variables that may be important when human actors are involved (e.g., the technique may lead to information overload or other types of complacency).
Moreover, solution evaluations that rely on historical data alone assume that future developers will use the new intervention in exactly the way the previous intervention was used, but this is not likely the case~\cite{LMP16}.

For {descriptive} research that aims to capture human and social aspects, data alone may also not tell the whole story and any conclusions drawn should be corroborated with other methods (such as interviews, surveys, or observations).
For example, Aranda and Venolia showed in their paper~\cite{Aranda2009} that many important aspects of software bugs cannot be discerned from data alone. 
Similarly, two papers that look at Git and GitHub as rich data sources (``Promises and Perils'' of Mining Git \cite{Bird2009} and GitHub \cite{Kalliamvakou2014}) highlight the many potential pitfalls when using these data sources alone.

\subsection{Why Are Human-Oriented Research Strategies Less Common in Software Engineering Research?}

We saw relatively few respondent, laboratory and field strategies in the ICSE and EMSE papers we analyzed (see Fig.~\ref{fig:smsresearchstrategies}). Furthermore, many of these were presented as a secondary strategy to a data strategy and were of limited scope: some were informally conducted or reported\footnote{For example, one EMSE paper we read reported a user study but did not indicate how many participants were involved, nor who the participants were.}). 
Although these strategies show the potential for advantages in terms of generalizability, realism, and control of human behaviour variables, these strategies are seldom used, either as a main or secondary strategy, even for research that claims to benefit human stakeholders.
We do not know if the reason for this lack of use may be due to fewer submissions of such papers, or if they are less likely to be accepted.  

One possible reason for the apparent low use of these strategies across the paper venues we analyzed may be a \textbf{lack of expertise} by software engineering researchers that publish in those venues.  
Strategies that study human behaviours require expertise that is not typically taught in a software engineering or computer science educational program (when compared to sociology and other social sciences). 

A possible reason for the very few lab and field studies may be due to \textbf{limited access} to developer participants and sites.
Field studies can be \textbf{obtrusive} and \textbf{practitioner time} is expensive. 
We found that only 27\% of the papers (29\% and 25\% of the EMSE and ICSE papers, respectively) reported involving practitioners in their empirical studies and these counts include several studies where just one or two practitioners were used in a small judgment study (see Table~\ref{table:practitioner}). 
\begin{table}
\caption{This table shows the number/proportion of papers that involve \textbf{practitioners} in reported empirical studies for the two venues and for both combined (all).}
\label{table:practitioner}
\begin{tabular}{lllllll} 
\toprule
Practitioners? & \multicolumn{2}{c}{All} & \multicolumn{2}{c}{ICSE} & \multicolumn{2}{c}{EMSE}  \\ 
                       & Count & Proportion      & Count & Proportion       & Count & Proportion        \\\midrule
True                   & 41    & 0.27            & 17    & 0.25             & 24    & 0.29              \\
False                  & 110   & 0.73            & 51    & 0.75             & 59    & 0.71              \\
\bottomrule
\end{tabular}
\end{table} 

Another possible cause for the low appearance of laboratory and field studies may be because of their potential \textbf{lack of generalizability}. 
 Reviewers can easily attack poor generalizability as a reason to quickly reject a paper, which may deter researchers who have tried to use such strategies~\cite{WilliamsThesis}.
     
 Field and lab strategies also often rely on the use of \textbf{qualitative methods} and should be evaluated using quite different \textbf{criteria} than those used to evaluate quantitative research. 
 Reviewers that expect to see threats to validity, such as external, internal, and construct, may find qualitative criteria, such as credibility and transferability, as unacceptable and unfamiliar due to a \textbf{different epistemological stance}~\cite{felderer2019evolution}.  

Respondent strategies were used more often than lab and field strategies in the papers we analyzed, but not as often as we had anticipated.
Respondent strategies are often seen as easier to implement for many researchers as they are done in settings of convenience. 
But conducting surveys and analyzing survey data bring other challenges: designing surveys normally takes several iterations, and recruiting survey respondents that are a good sample of the studied population is challenging.  
Furthermore, conducting open-ended surveys is \textbf{time consuming}\footnote{\url{http://www.gousios.gr/blog/Scaling-qualitative-research.html}}. 
 
  Finally, running studies with human subjects requires the additional step (in many academic institutions) of acquiring approval from an \textbf{ethics} review board~\cite{singer2002ethical}, and the use of human subjects inevitably introduces other complications (sometimes people do not show up or have unique abilities that impact the study).
 This step is generally not needed for data strategies, although ethical concerns about the use of some data resources have been raised in the research community.

\subsection{Using Triangulation to Balance Benefits for Human and Technical Aspects in Empirical Software Engineering Research}
\label{sec:triangulate}

Denzin \cite{Denzin1973} describes several different types of triangulation in research:
investigator triangulation, data triangulation and methodological triangulation.
Methodological triangulation refers to using different strategies and/or methods, while data triangulation refers to when the same research method may be used but different sources of data are used.
Investigator triangulation refers to the use of different investigators in running studies to reduce investigator bias.
Triangulation of research strategy (what Denzin refers to as methodological triangulation) is how researchers can improve the balance of desirable research quality criteria \cite{runkel1972}.
Our framework's circumplex (Fig. \ref{fig:framework}) highlights how each strategy has strengths and weaknesses when it comes to improving generalizability, enhanced control over variables that may be influenced by human actors, improved study realism, and more precision over data measurements.
The choice of research strategy, and more specifically which methods form part of that strategy, indirectly influences \textit{who} benefits from the research: practitioner stakeholders, technical components/systems or researchers.

The potential impact of the choices made by the authors of the papers we analyzed on four research quality criteria is summarized in Fig.~\ref{fig:smsresearchstrategies}. 
Most notably, few papers report on research strategies that attempt to control variables that come into play when human actors may be involved (e.g., very few report on lab and field experiments).  
Papers that do not control for human variables are limited in how they may claim their research is relevant to or benefits human actors.
Realism is also potentially low in the cases where a realistic evaluation should involve human participants when the research claims to benefit them.
In contrast, the use of data strategies may help improve generalizability when datasets from multiple cases are considered, and such papers potentially achieve higher precision over data measurements from the technical systems studied.

\begin{figure} [h]
	\centering
    \includegraphics[width=200pt]{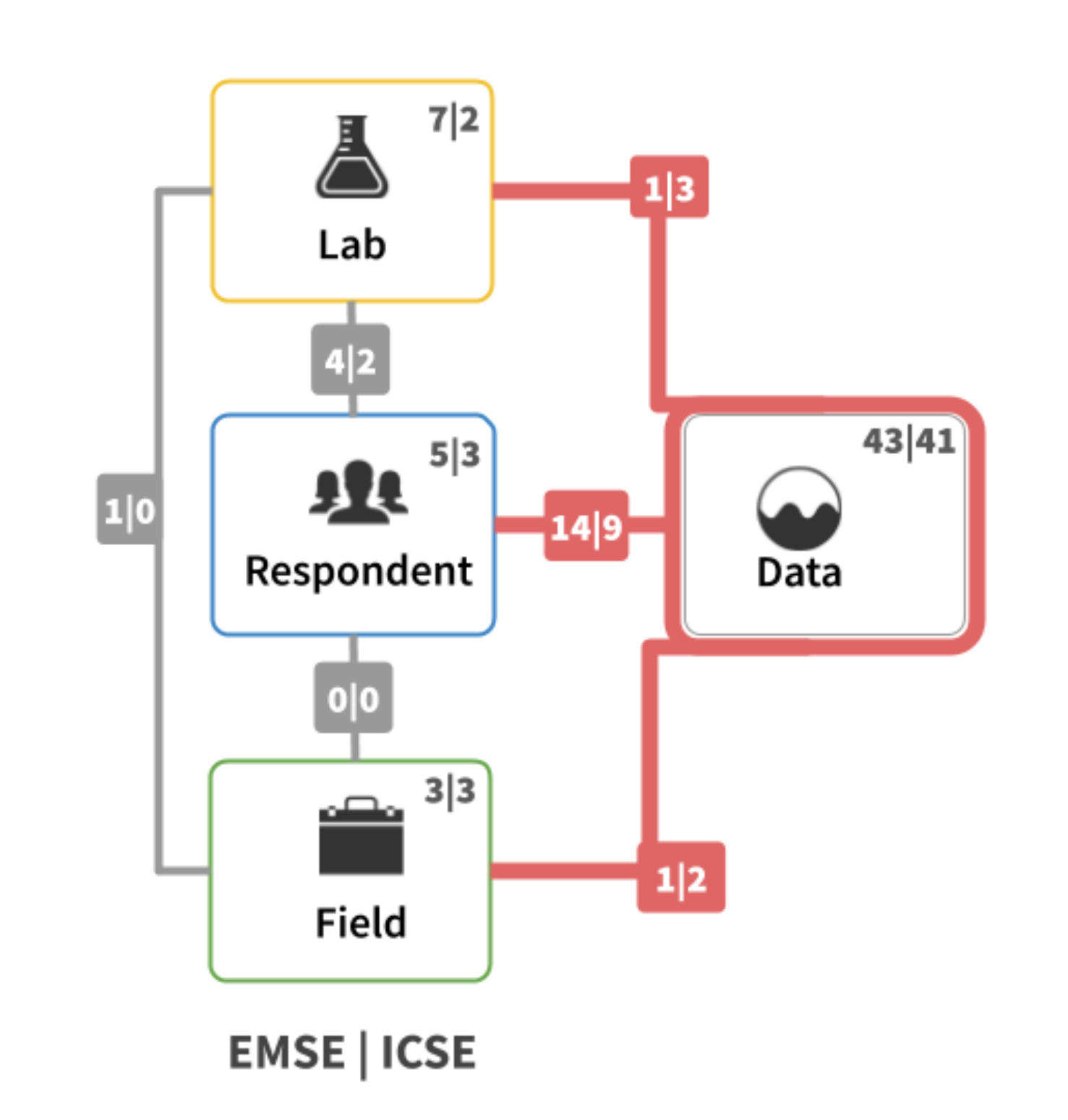}
    \caption{\textbf{Triangulation} of research strategies reported in our set of EMSE/ICSE papers, respectively. Numbers placed on the edges indicate the number of papers that triangulate using that pair of strategies. The red (thicker) edges indicate triangulation of data strategies with non-data strategies. The numbers in the white strategy boxes indicate number of strategies that were not triangulated with other strategies. }
    \label{fig:triangulation}
\end{figure}

In our study of the 
EMSE journal and ICSE conference papers from 2017, only 37 papers (24\%/25\% of the EMSE/ICSE papers respectively) reported studies from more than one research strategy quadrant.
Figure~\ref{fig:triangulation} shows which strategies were triangulated with one another.
For the data strategy papers, the vast majority did not triangulate their findings with other strategy quadrants, but when triangulation did occur, data strategies were mostly triangulated with a respondent strategy (surveys or interviews).  

 Although triangulation of research strategy was low within the papers we analyzed and across the venues we studied, we found that data triangulation was quite common in our study sample. 
Data triangulation occurred when a paper author replicated results with additional cases (datasets) using data-driven strategies. 
While the use of data triangulation may improve generalizability, it does not improve realism or control over human variables. 
These latter two criteria are particularly important to improve for research that is explicitly aimed at human beneficiaries.
 
In terms of diversifying strategies, there are other benefits and drawbacks to be considered (in addition to realism, control over human variable, generalizability and data precision). 
Field strategies often lead to immediate actionable descriptive insights and understanding of why things work the way they do, as well as to innovative solution ideas from watching how experienced and creative developers deal with or work around a problem in the real world. 
Likewise, non-adoption of a particular solution in the field can lead to innovative solutions.
On the other hand, data strategies may be easier to replicate, while both respondent and data strategies may more easily scale to larger and broader populations, potentially increasing generalizability.

We could not determine whether authors triangulated their work outside of the papers we considered, and we
recognize that the responsibility for triangulation does not need to be on the level of individual papers.
Given that EMSE is a journal and many papers extend conference papers, we expected to see \textbf{more triangulation} in EMSE papers, however, we did not see this in our analysis (see Fig. ~\ref{fig:triangulation}).
For ICSE papers, we see that data strategies are frequently used for solution papers but more diverse strategies are used for descriptive papers. 
For EMSE, we see that even descriptive papers rely more on data strategies alone.
Of course, the EMSE authors of descriptive papers may have relied on data strategies that use human-generated trace data to study human and social aspects, but in doing so, may have missed other context variables that limit the results.

Finally, the choice of research strategy only offers the \emph{potential} to achieve a given criterion. Many studies do not maximize this potential: they may use students instead of professional developers (reducing realism, see Table~\ref{table:practitioner}), have low statistical power (reducing generalizability), convenience sample from unrepresentative populations (reducing control), or make questionable analysis choices (reducing precision). Each method must still be judged on its merits and according to best practices.

In sum, multiple research strategies should be used to support triangulation, 
not just to triangulate specific findings, but to further add to insights concerning software engineering problem contexts, possible solutions for those problems, and the evaluation of those solutions~\cite{Engstrom2019}. 
Doing so will allow our research to be more relevant and transferable to broader problem contexts, leading to richer theories about why proposed solutions work or do not work as expected in certain human and social aspects.

\section{Limitations}
\label{Limitations}

We identify the limitations associated with this research and the measures we took to mitigate these issues through our research design. We use the Roller and Lavrakas~\cite{Roller2015} ``Total Quality Framework'' (TQF) for qualitative research, consisting of the subdomains ``Credibility" (of the data collection), ``Analyzability" (of data analysis), ``Transparency" (of the reporting), and ``Usefulness" (of the results). 
The TQF, being specifically derived from experiences in qualitative research, is more relevant to this paper than the typical ``internal/external/construct'' frame that is often applied in statistical studies. It closely parallels the discussions of high-quality qualitative research in Miles, Huberman and Saldana~\cite{mileshuberman2013}, Kirk and Miller~\cite{KM86}, and Onwuegbuzie and Leech~\cite{ol2007}, among others (a more complete discussion can be found in Lenberg \emph{et al.}~\cite{lenberg2017behavioral}).

\subsection{Credibility} 

Credibility is an assessment of the \textbf{completeness and accuracy} of the data gathering part of the study. 

The scope of our study was limited to full research papers from the EMSE journal and ICSE conference in 2017. 
Different years, different venues, and different tracks may have produced a different distribution of research beneficiary, research contribution, and research strategy use. 
However, we show our analysis as an example of the Who-What-How framework applied to two venues and do not claim this to be exhaustive. 
We know from an earlier iteration of a similar study that considered the preceding two years at ICSE (2015 and 2016), as well as ICSE 2017, that the trends were extremely similar~\cite{WilliamsThesis}. 
Journals and conferences are different venues, and in software engineering they are seen to serve different purposes so some differences were expected. However, EMSE and ICSE serve similar communities. For example, 26 of the 201 members of the 2020 ICSE Technical Program Committees and EMSE Editorial Board are in common so some similarities were anticipated and observed.
However, the Who-What-How framework did help illuminate some similarities and differences between the ICSE and EMSE venues.

As mentioned earlier, there are also other venues that clearly focus on human aspects, including the CHASE workshop that has been co-located with ICSE since 2008, as well as other venues such as VL/HCC\footnote{Visual Languages and Human-Centric Computing, \url{http://conferences.computer.org/VLHCC/}} and CSCW\footnote{ACM Conference on Computer Supported Cooperative Work \url{https://cscw.acm.org}}. Thus we recognize our findings are particular to the ICSE technical track and EMSE journal, but we feel it is important to share as these venues are recognized as being inclusive in terms of topics and methods, but are also seen by many as two of the premier publishing venues in software engineering. 

\subsection{Analyzability} 

Analyzability is an assessment of the accuracy of the paper's \textbf{analysis and interpretations}.

The authors come from a predominantly human-centered software research background and so this may have influenced our interpretation of papers outside our area of expertise (e.g., for areas such as automated testing). 
Our analysis tasks relied on human judgment in terms of classification of the beneficiary, contribution, and research strategy in the papers we analyzed. 

To consider who the \textbf{beneficiary} was, we relied on the paper text to discern if the research contribution of the paper aimed to benefit human stakeholders (e.g., developers, managers, end users), researchers (e.g., tool designers), or technical systems (e.g., a build system or recommender). If the study intended for other beneficiaries, or had other beneficiaries as some eventual outcome of a wider research program, our analysis would not be able to identify this.
Many papers alluded to multiple beneficiaries. In the case where this was somewhat subjective on our part, we included a quote from the paper in our spreadsheet---in particular, we included quotes when we identified the research was aimed at a human stakeholder beneficiary. 

We assigned each paper a single \textbf{contribution} type for the research study (i.e., either descriptive or solution). However, we recognize this is a coarse description of a single paper's epistemological goals. For example, some studies might do exploration leading to descriptive insights and then design or propose a tool. 
We relied mostly on the authors' own descriptions in their papers to help us decide if descriptive or solution was the best categorization for research contribution.
A richer categorization could leverage design science terminology more fully, such as in Engstr\"{o}m \emph{et al.}~\cite{Engstrom2019}, but such a detailed categorization was beyond the scope of our research. 

For coding \textbf{research strategy}, we relied on the strategy quadrants we identified in the \emph{how} part of the framework we developed in Section~\ref{Framework}. 
This aspect of our framework was developed with a focus on studying human aspects. 
In an earlier study reported by Williams (one of this paper's authors), she considered the research strategy across ICSE technical track papers from 2015-2017 and asked authors to classify their own papers according to the research strategy used~\cite{WilliamsThesis}.
She found that her categorization very closely aligned with the authors' views and any discrepancies were easily resolved (for example, an author may have been confused by the terminology we used or forgot that they used an additional strategy in their work).   
Although this was an earlier version of the research strategy (How) part of our framework, this earlier finding increases our confidence that our assignment of research strategy would match the authors' views at least for the ICSE 2017 papers. 
We note, however, that different frameworks for categorizing research strategy would lead to different categorizations (e.g., see a related framework by Stol \emph{et al.}~\cite{Stol2018}).
 
To mitigate researcher bias for all of the coding we did, two of us independently coded the papers. We revisited any codes we disagreed on, and reread the papers together to arrive at an agreement. For some papers, we recruited additional readers to join our discussions.
We recognize that some of our codes may still be open to interpretation, and thus we make our 
spreadsheet available for others to peruse and to validate our analysis.  
Our spreadsheet contains many quotes from the papers we analyzed to help others understand our coding decisions.

\subsection{Transparency}
Transparency refers to the way in which we have presented our results. We rely mainly on the replication package to ensure transparency. 
Our coding spreadsheet has additional information on any of the papers where we disagreed or where we felt our coding may have been subjective. We capture this in comments associated with each paper in the spreadsheet. 
We can also rely, to some extent, on the familiarity readers likely have with the domain. Since we write about software research, we did not see a need to provide detailed descriptions of the domain. 

\subsection{Usefulness}
We developed and presented a socio-technical research framework---the Who-What-How framework---for reflecting on socio-technical contributions in software engineering research.
We believe our framework is ready to apply to other venues (e.g., other software engineering journals, conferences, or tracks).
To facilitate this further application, we provide a number of documents on our supplementary website designed to help other researchers follow our methodology.
We welcome replication studies---especially triangulation studies with new research strategies---on additional years of ICSE or EMSE and other venues to explore the differences that may exist between venues and time periods in software engineering research.
Finally, although we do not demonstrate this in our paper, we have found anecdotally from our students and colleagues that the framework is useful in helping design and reflect on research.

\section{Conclusions}
\label{Conclusion}

Understanding the complexities of human behavior requires the use of diverse research strategies---specifically the use of strategies focused on human and social aspects.  
Through our analysis of 151 ICSE technical track papers and EMSE journal papers from 2017, we found a skew towards data strategies in these publishing venues, even for papers that claimed potential benefits for human stakeholders, in addition to their claimed improvements to technical components.
Relying on data strategies alone may mean we miss important aspects of the complex, socio-technical context of software engineering problems and
 hinder our evaluation of how tools may be used in real practice scenarios.

We might expect initial research on a socio-technical software engineering problem to consider only technical aspects or rely only on limited behavioral trace data. 
But at a community level, we would expect to see more studies that expand on these initial works to rigorously examine the human aspects. 
In the cohort of papers we analyzed, we found that a minority of data strategy papers triangulated using additional strategies, many of which were limited in scope, while the majority did not triangulate their research strategies at all.

Earlier efforts on analyzing research at particular publishing venues focused on categorizing research and empirical studies in software engineering (see Section~\ref{Background}), while the work we report in this paper focuses specifically on how studies approach social aspects, and discusses the trade-offs and implications for the software engineering community's collective knowledge on the choices made in the studies we examined.
We encourage other researchers to use the framework and apply it to reflect on other publishing venues, and possibly compare with the EMSE journal and ICSE technical track venue analysis we report. 

Although our original intention was to use this framework to reflect on human and social aspects of existing software engineering papers, 
we hope that the Who-What-How framework is also useful for framing or designing new or in-progress research studies, and to reflect on the implications of one's personal choice of strategies in terms of generalizability, realism, precision over data and control of human aspects.
 
Finally, we hope the Who-What-How framework will lead to some community-wide discussions about the overall shape of the research we do, and how certain values and expectations at the level of a particular publishing venue may impact the relevance of our research for researchers and practitioners.

\begin{acknowledgements}
We would like to thank Cassandra Petrachenko, Alexey Zagalsky and Soroush Yousefi for their invaluable help with this paper and research. We also thank Marian Petre and the anonymous reviewers for their insightful suggestions to improve our paper.
\end{acknowledgements}

\bibliographystyle{spmpsci}
\bibliographystyleicse{spmpsci}      
\bibliography{emse2019}   \bibliographyicse{icserefs}

\appendix

\appendix 

\section{The Circumplex of Runkel and McGrath}
\label{AppendixA}

\begin{figure}[htb]
    \includegraphics[width=\linewidth]{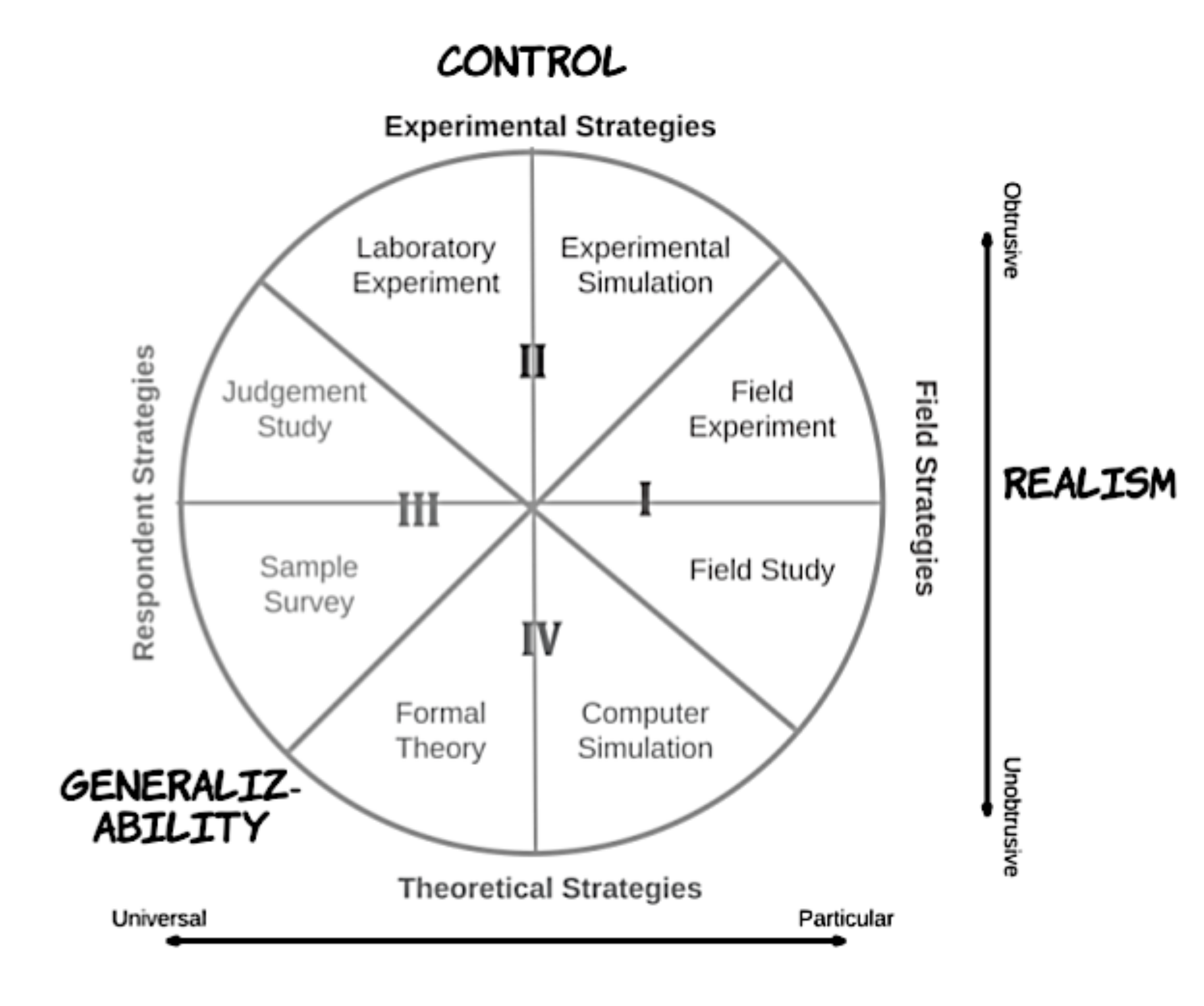}
    \caption{Runkel and McGrath's research strategy circumplex.}
    \label{fig:mcgrath}
\end{figure}

Figure~\ref{fig:mcgrath} shows a sketch of the research strategy circumplex designed by Runkel and McGrath \cite{runkel1972} for categorizing behavioral research strategies. We adapted their model for the \emph{How} part of our research framework. 
Runkel and McGrath's model of research strategies was developed in the 1970s for categorizing human behavioral research, hence it provides a good model for examining socio-technical factors in software engineering. 

The McGrath model has been used by other software engineering researchers to reflect on research strategy choice and its implications on research design~\cite{Easterbrook2008}, and most recently by Stol and Fitzgerald~\cite{Stol2018} as a way to 
 to provide consistent terminology for research strategies~\cite{Stol2018}~\footnote{Stol and Fitzgerald interpret and extend this model quite differently to us as they are not concerned with using their framework to discriminate which strategies directly involve human actors.  Runkel and McGrath developed their model to capture behavioral aspects and we maintain the behavioral aspect in our extension of their model.}
It is used extensively in the field of Human Computer Interaction~\cite{Baecker95} and CSCW~\cite{cruz2012towards} to guide research design on human aspects. 

Three of our quadrants (Respondent, Lab, Field) mirror three of the quadrants in Runkel and McGrath's book (although we refer to Experimental Strategies as Lab Strategies as we find this less confusing).   
The fourth quadrant they suggest captures non-empirical research methods: they refer to this quadrant as Theoretical Strategies.
We consider two types of non-empirical strategies in our framework:
 Meta (e.g., systematic literature review), and Formal Theory. 
 We show these non empirical strategies separately to the four quadrants of empirical strategies in our framework.  
 Our fourth quadrant includes Computer Simulations (which we consider empirical), but it also includes other types of data strategies that rely solely on previously collected data in addition to simulated data. 
 We call this fourth quadrant in our framework ``Data Strategies''.

One of the core contributions of the Runkel and McGrath research strategy model is to highlight the trade-offs inherent in choosing a research strategy and how each strategy has strengths and weaknesses in terms of achieving higher levels of generalizability, realism and control.
Runkel and McGrath refer to these criteria as ``quality criteria'', since achieving higher levels of these criteria is desirable. 
\emph{Generalizability} captures how generalizable the findings may be to the population outside of the specific actors under study. 
\emph{Realism} captures how closely the context under which evidence is gathered may match real life. 
\emph{Control} refers to the control over the measurement of variables that may be relevant when human behaviors are studied.
Field strategies typically exhibit low generalizability, but have higher potential for higher realism.
Lab studies have high control over human variables, but lower realism. 
Respondent strategies show higher potential for generalizability, but lower realism and control over human variables. 

We added a fourth research quality criterion to our model, data \emph{precision}.
Data strategies have higher potential for collecting precise measurements of system data over other strategies. 
Data studies may be reported as `controlled' by some authors when they really mean precision over data collected, therefore, we reserve the term control in this paper for control over variables in the data generation process (e.g., applying a treatment to one of two groups and observing effects on a dependent variable). 
McGrath himself debated the distinction between precision and control in his later work. 
We note that McGrath's observations were based on work in sociology and less likely to involve large data studies, unlike in software engineering.
The Who-What-How framework (bottom of Fig. \ref{fig:framework}) denotes these criteria in italics outside the quadrants. The closer a quadrant to the criterion, the more the quadrant has the  potential to maximize that criterion.

We recommend that the interested reader refer to Runkel and McGrath's landmark book \cite{runkel1972} for additional insights on methodology choice that we could not include in our paper.

\section{Sample Paper Classification}
\label{Appendix B}
Table \ref{tbl:samplesheet} shows a 15-paper sample classified using our Who-What-How framework. Full data is available at \replicationUrl.
    \thispagestyle{empty}\begin{sidewaystable}
\begin{tabular}{cp{5cm}lcccc}
\toprule

Venue & Paper Title & Authors & Strategies & Purpose & Beneficiaries \\ \midrule
ICSE & The Evolution of Continuous Experimentation in Software Product Development & Fabijan \emph{et al.} & FS & Descriptive & Human \\
ICSE & Learning Syntactic Program Transformations from Examples & Rolim \emph{et al.} & D & Solution & Human\_System \\
ICSE & Glacier: Transitive Class Immutability for Java & Coblenz \emph{et al.} & D/LE & Solution & Human\_System \\
ICSE & UML Diagram Refinement & Faitelson & FT  & Solution & Human \\
ICSE & Understanding the Impressions, Motivations and Barriers of One Time Code Contributors to FLOSS Projects: A Survey & Lee \emph{et al.} & SS  & Descriptive & Human \\
ICSE & Fuzzy Fine-grained Code-history Analysis & Servant \emph{et al.} & D/JS & Solution & Human\_System \\
ICSE & On Cross-stack Configuration Errors & Sayagh \emph{et al.} & D/D & Solution & All \\
ICSE & Machine Learning-Based Detection of Open Source License Exceptions & Vendome \emph{et al.} & D/JS & Descriptive & Human\_System \\
ICSE & Feedback-Based Debugging & Lin \emph{et al.} & D/LE & Solution & Human\_System \\
EMSE & On the long-term use of visual GUI testing in industrial practice - a case study. & Alégroth \emph{et al.} & FS & Descriptive & Human \\
EMSE & Evaluating code complexity triggers, use of complexity measures and the influence of code complexity on maintenance time. & Antinyan \emph{et al.} & SS & Descriptive & Human \\
EMSE & Reengineering legacy applications into software product lines - a systematic mapping. & Assunçao \emph{et al.} & Meta  & Descriptive & Researcher \\
EMSE & User satisfaction and system success - an empirical exploration of user involvement in software development. & Bano \emph{et al.} & FS/Meta & Descriptive & Human \\
EMSE & Extracting and analyzing time-series HCI data from screen-captured task videos. & Bao \emph{et al.} & D/LE & Solution & System\_Researcher \\
EMSE & The last line effect explained. & Beller \emph{et al.} & D/SS & Descriptive & Human\_System \\ \bottomrule
\end{tabular}
	\caption{Examples of our paper classification and coding. FS: Field Study, D: Data Study, LE: Lab Experiment, JS: Judgment Study, FT: Formal Theory, SS: Sample Study.}
	\label{tbl:samplesheet}
\end{sidewaystable}

\end{document}